\documentclass[aps,prb,twocolumn,amsmath,amssymb]{revtex4}

\usepackage{epsfig}
\usepackage{graphicx}
\usepackage{amsfonts}
\usepackage{amssymb}
\usepackage{amsmath}
\usepackage{amsfonts}
\usepackage{bm}
\usepackage{bbold}
\usepackage{color}
\usepackage{float}

\renewcommand{\vec}[1]{{\mathbf #1}}

\bibpunct{(}{)}{;}{s}{,}{,}

\def\be{\begin{equation}}
\def\ee{\end{equation}}
\def\bea{\begin{eqnarray}}
\def\eea{\end{eqnarray}}
\begin{document}

\title{Fulde-Ferrell States in Inverse Proximity Coupled Magnetically-Doped Topological Heterostructures}

\author{Moon Jip Park$^{1,2}$}
\author{Junyoung Yang$^{1}$}
\author{Youngseok Kim$^{2,3}$}
\author{Matthew J. Gilbert$^{2,3}$}
\affiliation{$^1$ Department of Physics, University of Illinois, Urbana, IL 61801}
\affiliation{$^2$ Micro and Nanotechnology Laboratory, University of Illinois, Urbana, IL 61801}
\affiliation{$^3$ Department of Electrical and Computer Engineering, University of Illinois, Urbana, IL 61801}

\date{\today}

\begin{abstract}
We study the superconducting properties of the thin film BCS superconductor proximity coupled to a magnetically doped topological insulator(TI). Using the mean field theory, we show that Fulde-Ferrell(FF) pairing can be induced in the conventional superconductor by having inverse proximity effect(IPE). This occurs when the IPE of the TI to the superconductor is large enough that the normal band of the superconductor possesses a proximity induced spin-orbit coupling and magnetization. We find that the energetics of the different pairings are dependent on the coupling strength between the TI and the BCS superconductor and the thickness of the superconductor film. As the thickness of the superconductor film is increased, we find a crossover from the FF pairing to the BCS pairing phase. This is a consequence of the increased number of the superconducting bands, which favor the BCS pairing, implying that the FF phase can only be observed in the thin-film limit. In addition, we also propose transport experiments that show distinct signatures of the FF phase.
\end{abstract}
\pacs{71.35.-y, 73.20.-r, 73.22.Gk, 73.43.-f}
\maketitle

\section{introduction}
The surface state of the time reversal invariant topological insulator(TI) exhibits a spin momentum locked massless Dirac fermion\cite{Zhang2009,Zhang2010,doi:10.1021/nl200584f,PhysRevLett.98.106803}.
This spin-orbit coupled band has been explored for potential applications to spintronics\cite{Baker2015,PhysRevB.93.125303,Mellnik2014,PhysRevLett.109.166602} and topological quantum
computation\cite{PhysRevLett.100.096407,PhysRevB.86.155431,PhysRevLett.102.216404,PhysRevB.84.201105,Li2014}. Among these various applications, the superconducting surface state has gathered a lot of interest due to the possible emergence of the Majorana fermions when proximity coupled to BCS superconductor\cite{PhysRevLett.100.096407,PhysRevB.94.235434,PhysRevB.86.155431,PhysRevLett.102.216404,PhysRevB.84.201105,Li2014}.
The underlying idea of this proximity coupled system is that resulting system mimics a $p+ip$ superconductor\cite{PhysRevLett.100.096407}. Along these lines, it has been proposed to utilize the proximity coupled surface states as a new platform to realize unconventional superconducting states  with the non-zero spin and the angular momentum\cite{Xu2014,PhysRevB.87.220506,PhysRevB.92.205424}. The proposed unconventional superconducting states include the helical pairing state with $p\pm ip$ pairing\cite{Xu2014}, and the spin triplet pairing in the magnetically doped TI\cite{PhysRevB.92.205424}, and the odd frequency superconductivity in the thin film of the TI\cite{PhysRevB.90.184517}.

Besides the non-trivial angular momentum states of the Cooper pairs, the superconducting states with the non-trivial linear momentum have been proposed to exist in strong spin-orbit coupled materials other than TI\cite{Zheng2014,1367-2630-15-9-093037,PhysRevA.87.031602,PhysRevA.87.031603,PhysRevB.86.214514,PhysRevB.89.014506,PhysRevB.92.035153,PhysRevB.93.094517,PhysRevB.93.214511}. The examples are spin orbit coupled Fermi gases in the cold atom systems\cite{Zheng2014,1367-2630-15-9-093037,PhysRevA.87.031602,PhysRevA.87.031603} and the bulk doped Weyl semimetals\cite{PhysRevB.86.214514,PhysRevB.89.014506,PhysRevB.92.035153,PhysRevB.93.094517,PhysRevB.93.214511}. This unconventional superconducting state is known as Fulde-Ferrell-Larkin-Ovchinnikov(FFLO) phase, whose Cooper pairs in equilibrium have a non-zero linear momentum. The FFLO phase is predicted to exhibit phenomena, which is not found in the conventional BCS superconductivity such as a spatial modulation of the pairing potential\cite{PhysRev.135.A550,larkin:1964zz} in equilibrium. The FFLO phase has been proposed to exist without the external magnetic flux in the context of the superconductivity when paired with the ferromagnetic alloys\cite{PhysRev.135.A550} which utilize spin imbalanced Fermi surface to generate finite momentum pairing\cite{Liao2010}. Interestingly, a non-trivial Fraunhofer pattern in HgTe quantum well\cite{Hart2016}, which may support the finite momentum pairing, has been recently observed with the application of the external magnetic field. Nevertheless, a clear signature of the FFLO pairing in the absence of the external field is lacking in the condensed matter systems. While the spin-orbit coupled materials are known to offer a larger parameter space to support the FFLO phase\cite{Zheng2014}, it is desirable to explore candidate spin-orbit coupled systems comprised of readily available materials. In this regard, we propose the ground state with the FF pairing can occur in a conventional BCS superconductor that is proximity-coupled to a magnetically doped TI surface state. The FF pairing is a specific type of the FFLO phase that has spatial modulation of the order parameter phase while the LO pairing has the modulation of the amplitude. In our setup, the magnetic dopants within the topological insulator induce a uniform Zeeman field pointing in a direction parallel to the surface. Our proposal has clear advantages in the experimental accessibility: i) The proximity coupled superconductivity on the surface of the topological insulator has been widely realized\cite{PhysRevLett.114.017001,Zareapour2012,Xu2014,PhysRevB.86.134504,PhysRevB.90.085128,PhysRevLett.114.017001,PhysRevB.84.165120,Wang52,doi:10.1021/nl400997k,Cho2013,1367-2630-16-12-123043}. ii) The magnetic energy gap in the magnetically doped surface of TI with non-zero exchange field has been observed\cite{Chang167,Yu61,Chen659}. iii) There is no magnetism inside the superconductor, which might destroy the superconductivity.

In this work, we analyze the energetics of a proximity-coupled magnetically doped TI-superconductor structure to determine the stability of the FF phase as function of experimentally relevant parameters. First, in section \ref{model}, we introduce a model that describes the proximity coupled structure of the topological insulator and the conventional superconductor, which utilizes the low energy model of the bands of $Bi_2Se_3$ derived from ARPES experiment and $NbSe_2$ derived from first principle calculation to accurately capture the relevant physics of the recent experiments\cite{PhysRevLett.114.017001,Xu2014,1367-2630-16-12-123043}. In this model, we choose the model of the superconductor to be $NbSe_2$ due to its wide use in experiments\cite{Xu2014}. In section \ref{inverseproximity}, we show that the metallic band of the superconductor exhibits an anisotropic Fermi surface in the Brillouin zone as a consequence of the 'inverse proximity effect'(IPE), which we denotes as the proximity effect of the topological insulator acting upon the superconductor\cite{Shoman2015}. In section \ref{numerics}, we use mean-field theory of superconductivity to calculate the energetic stability of the BCS pairing and the FF pairing. We show that the  metallic band of the superconductor can have a FF pairing as its ground state in the thin film limit due to the proximity induced anisotropic Fermi surface, where the decaying length of the IPE exceeds the entire region of the superconductor. We also consider the case where the thickness of the superconductor exceeds the penetration length of the IPE. We find that the FF pairing becomes unstable as the thickness of the superconductor increases, since more BCS favored bands become populated and overwhelm the FF phases. Nevertheless, we show that the FF pairing can survive at the interface of the heterostructure in the thick sample limit. In section \ref{transport}, we propose two transport methods. Our transport setups show distinct transport signatures that distinguish the FF phase from that of the conventional Josephson junction. Finally, in section \ref{conclusion}, we conclude our study and summarize our results. 

\section{model} \label{model}
\newcommand{\updownarrows}{\mathbin\uparrow\hspace{-.0em}\downarrow}
\newcommand{\downuparrows}{\mathbin\downarrow\hspace{-.0em}\uparrow}
\newcommand{\up}{\uparrow}
\newcommand{\dn}{\downarrow}
\newcommand{\braket}[1]{\langle #1 \rangle}

\begin{figure}
\includegraphics[width=0.5\textwidth]{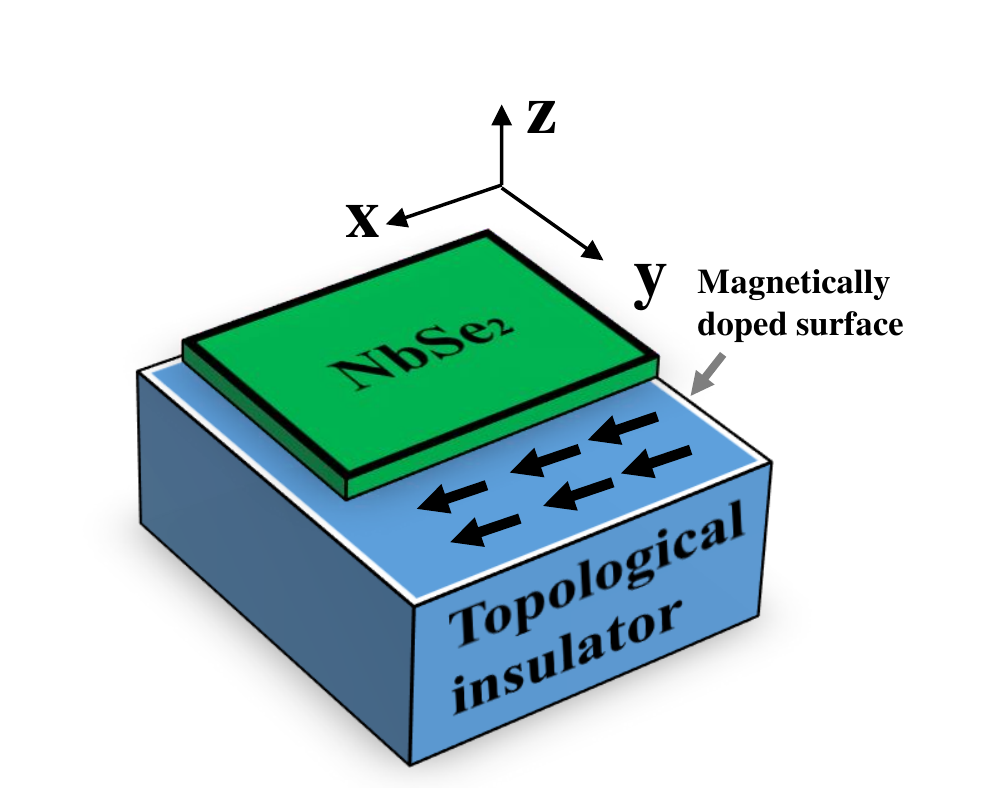}
\caption{\label{schematic}
The schematic figure of the magnetically doped topological insulator superconductor hetero-structure. On the top of the topological insulator surface, a thin film of the BCS superconductor is deposited. The magnetization points out the parallel direction to the surface of the topological insulator to shift the location of the Dirac cone.}
\end{figure}

In Fig. \ref{schematic}, we show the system comprised of a metallic superconductor grown on top of a magnetically doped 3D TI. We choose the NbSe2 as our metallic superconductor as it has been widely used for the TI-superconductor heterostructure due to its lattice matching with the $Bi_2Se_3$. We begin our discussion by writing down the metallic Hamiltonian which describes the parent superconductor of $NbSe_2$: \cite{2008Inosov,2012Rahn}
\begin{equation} \label{eq:Hm}
\hat{H}_{M}(\vec{k})=t_M k^2-\mu_M
\end{equation}
where  $k=\sqrt{k_x^2+k_y^2}$ is the magnitude of the in-plane momentum, $t_M$ is the material parameter that determines the slope of the parabolic band. $a$ is the lattice constant, and $\mu_M$ is the chemical potential. We choose the value of these parameters to be $a=0.344nm$,$t_M=-0.5eV/a^2$, and $\mu_M=0.8eV$. $\mu_M$ is obtained by fitting the tight-binding Hamiltonian\cite{2008Inosov,2012Rahn} of the 2H-NbSe$_2$ to the quadratic band near the chosen chemical potential.
We now consider the surface state Hamiltonian of the magnetically doped 3D TI:
\bea \label{eq:HTI}
\hat{H}_{TI}(\vec{k})= v_F( k_x\sigma_y-k_y\sigma_x)+m\sigma_x -\mu_{TI} I_2,
\eea
where, without loss of generality, we set $\hbar=1$.
$v_F$ is the Fermi velocity of the TI surface state, $m$ is the exchange field Zeeman term, $\mu_{TI}$ is the chemical potential of the topological insulator, $I_2$ is $2\times2$ identity matrix, and $\sigma_i$ is the $i$-th Pauli matrix for spin. The choice of the above parameters is taken from ARPES experiments\cite{PhysRevB.82.045122} of the surface bands to derive the values of the parameters: $v_F=1.19eV$ and $\mu_{TI}=0.26$. From Eqs. (\ref{eq:Hm}) and (\ref{eq:HTI}), our system is described by the total Hamiltonian written as
\bea \label{eq:Htot}
H_{M-TI}=H_{M}+H_{TI}+H_{coupling}.
\eea
In Eq. (\ref{eq:Htot}), the metallic Hamiltonian is $H_{M}=\sum_\vec{k}\psi_{M,\vec{k}}^\dagger \hat{H}_M(\vec{k})\psi_{M,\vec{k}}$ where we define the 2 component spinor $\psi_{M,\vec{k}}=[d_{\vec{k}\up}, d_{\vec{k}\dn}]^T$, and $d_{\mathbf{k}\up}^\dagger$ ($d_{\mathbf{k}\dn}$) is up-spin (down-spin) electron creation (annihilation) operator of the metal. Likewise, the TI Hamiltonian is $H_{TI}=\sum_\vec{k}\psi_{TI,\vec{k}}^\dagger \hat{H}_{TI}(\vec{k})\psi_{TI,\vec{k}}$ where we define $\psi_{TI,\vec{k}}=[c_{\mathbf{k}\uparrow}, c_{\mathbf{k}\downarrow}]^T$ where $c_{k\uparrow}^\dagger$ ($c_{\mathbf{k}\downarrow}$) is up-spin (down-spin) electron creation (annihilation) operator.
In Eq. (\ref{eq:Htot}), we introduce $H_{coupling}$ which couples the TI and the metallic system as
\bea \label{eq:Htc}
H_{coupling}=\sum_{\mathbf{k},s=\updownarrows} t_c( c^\dagger_{\mathbf{k}s} d_{\mathbf{k}s} + d^\dagger_{\mathbf{k}s} c_{\mathbf{k}s} )
\eea
where $t_c$ is a coupling constant.
From the Hamiltonian in Eqs. (\ref{eq:Hm}, \ref{eq:HTI}, \ref{eq:Htc}), we construct the matrix form of the metal-TI Hamiltonian, $H_{M-TI}=\sum_\vec{k}\Psi_\vec{k}^\dagger \hat{H}_{M-TI}(\vec{k}) \Psi_\vec{k}$, where
\bea \label{eq:HMTI}
\hat{H}_{M-TI}(\vec{k})=
\begin{pmatrix}
\hat{H}_{M}(\vec{k}) &\hat{H}_{coupling} \\
\hat{H}^\dagger_{coupling} & \hat{H}_{TI}(\vec{k})
\end{pmatrix},
\eea
with the operator $\Psi_\vec{k}=[\psi_{TI,\vec{k}},\psi_{M,\vec{k}}]^T$, and the coupling Hamiltonian $\hat{H}_{coupling}=t_cI_2$.

\section{Inverse proximity effect} \label{inverseproximity}

\begin{figure*} 
\includegraphics[width=1\textwidth]{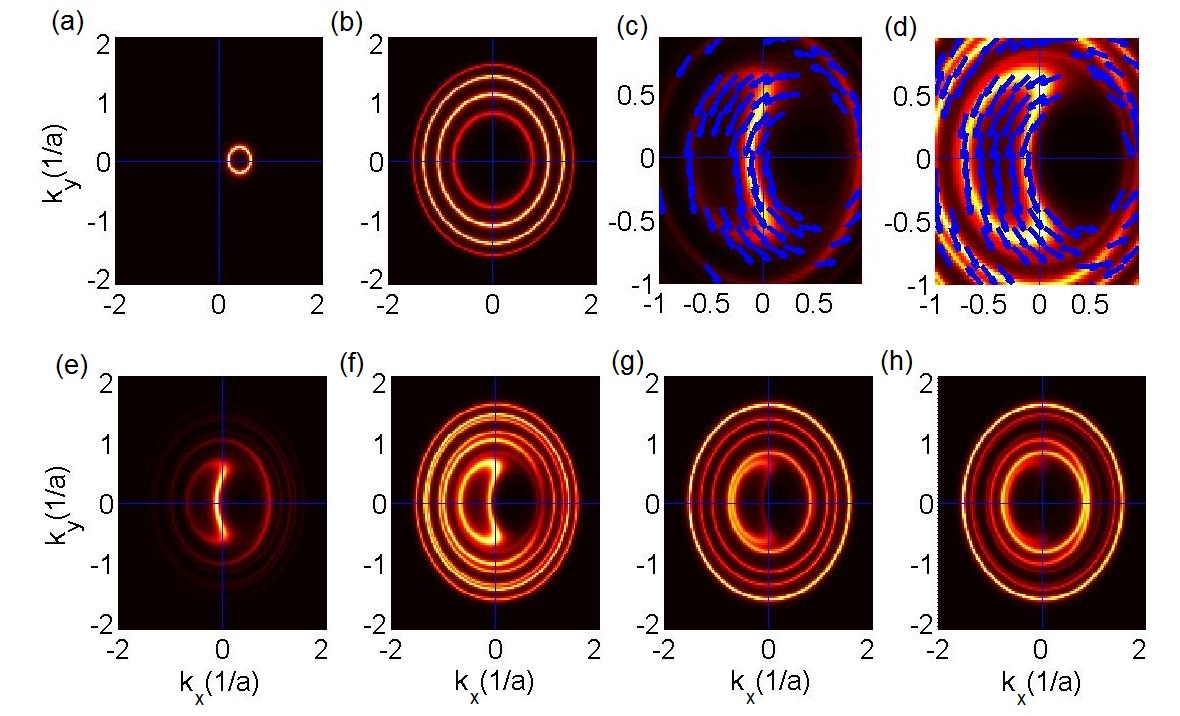}
\caption{\label{DOS} The local density of the state in momentum space at the topological insulator and the superconductor with $m=0.4eV$ and $t_c=0.5eV$. (a) and (b) show the LDOS of the topological insulator and the first layer of the superconductor when $t_c=0$. The magnetically ordered dopants shift the Dirac cone along the x-direction because the magnetism is aligned in the x-direction. As $t_c$ is turned on, the IPE starts to hybridize the Fermi surface. (c) and (d) show the LDOS and the spin texture. We find the non-zero spin orbit coupling and the effective Zeeman field in the superconductor. (e)-(h) show the LDOS of the topological insulator surface and the first, the second, and the third layer of the superconductor. We find that the Fermi surface starts to recover the isotropy and the IPE decays as we look further away from the surface of the topological insulator }
\end{figure*}
After establishing the description of the model and the Hamiltonian, we now consider the IPE of the TI to the metallic band of the superconductor. We choose to examine the IPE in this model in recognition of the fact that the IPE on the superconductor will significantly alter the metallic band and the corresponding superconductivity. The IPE can be evaluated by calculating the effective Hamiltonian of the metallic region in the presence of the finite coupling, $t_c$, with the TI. To calculate the effective Hamiltonian, we first consider the full matrix form of the Schrodinger equation without the superconducting pairing term, as given in Eq. (\ref{eq:HMTI}):
\bea
\label{fullsch}
\left(\begin{matrix}
      \hat{H}_{M} & \hat{H}_{couple} &  \\
       \hat{H}_{couple}^\dagger & \hat{H}_{TI} &   \\
   \end{matrix}\right)
\left(\begin{matrix}
      \psi_m\\
      \psi_{TI} \\
   \end{matrix}\right)
=E
\left(\begin{matrix}
      \psi_m\\
      \psi_{TI} \\
   \end{matrix}\right)
\eea
where $\psi_{m}$ and $\psi_{TI}$ is the wave function in the metallic and the TI region respectively and $E$ is the corresponding energy eigenvalue. To derive the self-energy term, $\hat\Sigma(E)$, which takes account the effect of the IPE, we integrate out the wave function in the TI region. The new effective Hamiltonian of the metal with the self-energy term now satisfies the following Schrodinger equation,
\bea
\hat{H}_{eff}(E)\psi_m=(\hat{H}_{m}+\hat{\Sigma}(E))\psi_m=E\psi_m
\eea
where the self energy is evaluated as,
\begin{gather}
\label{self}
\hat{\Sigma}(E)=\hat{H}_{couple}(E-\hat{H}_{TI})^{-1} \hat{H}_{couple}^\dagger
\\
\nonumber=\frac{t_c^2}{m^2+(v_F|k|)^2-\mu_{TI}^2}
\left(\begin{matrix}
      \mu_{TI}+m & -v_F(k_y +ik_x)  \\
       -v_F(k_y -ik_x) & \mu_{TI}-m &   \\
   \end{matrix}\right).
\end{gather}
As it has been shown from Eq. (\ref{self}), the effective metallic Hamiltonian now possesses a non-zero spin-orbit coupling and a non-zero Zeeman field with the coefficients of the strengths being $\frac{t_c^2v_F}{m^2+(v_F|k|)^2-\mu_{TI}^2}$ and $\frac{t_c^2m}{m^2+(v_F|k|)^2-\mu_{TI}^2}$, respectively. The presence of the both spin-orbit coupling and the Zeeman field distort the isotropic Fermi surface and shift the center of the momentum, which eventually decreases the energy gain from the Fermi surface instability under the singlet BCS pairing of the metal.

After deriving the analytical insight of the IPE, we now confirm the IPE by numerically evaluating the local density of state in the metallic region. The local density of state(LDOS) can be computed from the calculation of the imaginary part of the spectral function which is given as,
\begin{gather}
LDOS(\omega,i,k)=\sum_{n} -Im( \frac{|\phi_{n,i}(k)|^2}{\omega-E_n+i\eta})
\end{gather}
where $\eta$ is the infinitesimal broadening of the states, $\phi_{n,i}$ is the $n$-th eigenstate, and $E_n$ is the corresponding energy eigenvalue of the system. $i$ is the orbital degree of the freedom which represent the $z$ coordinate. 
The local spin density of state(LSDOS) can be similarly calculated by inserting pauli matrix by following
\begin{gather}
LSDOS_j(\omega,i,k)=\sum_{n} -Im( \frac{<\phi_{n,i}(k)|\sigma_j|\phi_{n,i}(k)>^2}{\omega-E_n+i\eta})
\end{gather}
In Fig. \ref{DOS}, we plot the numerically computed $LDOS(\omega,k)$ to show the change of the metallic band of the superconductor due to the IPE. Figs. \ref{DOS} (a) and (b) show the LDOS of the TI and the metallic layers, respectively, at $t_c=0$. In Fig. \ref{DOS} (a), we find that the Fermi surface of the TI is shifted in $\hat{x}$ direction as the finite Zeeman term shifts the location of the Dirac cone to $\Delta k_x=\frac{m}{v_F}$ in the Brillouin zone. Since the IPE($t_c=0$) is zero, Fig. \ref{DOS} (b) still shows the isotropic fermi surfaces of the metallic band in which one can always find a conventional BCS cooper pair with opposite momenta $\vec{K}$ and $-\vec{K}$ on the Fermi surface. On the other hand, as $t_c$ is turned on, we find that the the surface band of the TI and the metallic band of the superconductor layers start to hybridize. Fig. \ref{DOS} (c) and (d) shows the hybrdized Fermi surface and its spin texture of the TI and the first layer of the metal respectively. While the singlet superconducting pairing only couples the opposite spins, the LDOS depicted in Fig. \ref{DOS} (d) does not possesses a pair of the states that have opposite spins and zero net momentum simultaneously. On the other hand, Fig. \ref{DOS} (d) alternatively shows that a pair of the states with opposite spins have rather a finite net momentum along $x$ direction, which leads to the FF instability. In this case, it is not guaranteed to find two arbitrary electrons with the opposite momenta and the opposite spin on the Fermi surface. As a consequence, the BCS pairing may not be efficiently formed to lower the total ground state energy and, consequently, the finite momentum pairing phase may have lower ground state energy.
Indeed, we observe that the system with induced anisotropic Fermi surface favors FF states in certain parameter space in section \ref{numerics}. Additionally, Figs. \ref{DOS} (e)-(h) show the LDOS of the TI and the first, the second, and the third superconductor layer respectively. While Figs. \ref{DOS} (e) and (f) shows the same anisotropic LDOS shown in Figs. \ref{DOS} (c) and (d), we immediately observe that Fig. \ref{DOS} (g) and (h) show the anisotropy of the Fermi surface decays as we look further away from the interface between the metal and the TI. This is the consequence of the exponential decaying of the IPE away from the interface. As can be seen from Eq. (\ref{self}), the strength of the IPE decays exponentially as the inter-layer directional hopping of the $NbSe_2$ is known to be very small compared to the hopping in the intra-layer direction\cite{2008Inosov,2012Rahn}, so that we expect the proximity effect of the TI only survives in the first few layers. In the next section \ref{numerics}, we discuss the effect of the decaying IPE and the superconducting instability in the thick sample.

\section{Numerical Calculation of the ground state energies} \label{numerics}
\subsection{Single-layer limit of the superconductor}

With our understanding on the IPE of the metallic Hamiltonian, we consider superconducting phase to calculate ground state energy of the BCS and the FF states.
We first consider the $s$-wave superconducting order in metallic system. The $s$-wave pairing Hamiltonian in mean-field level is
\bea \label{eq:Hbcs}
H_{BCS}= -U\sum_\mathbf{k} [\Delta d^\dagger_{\mathbf{k}\up} d^\dagger_{-\mathbf{k}\dn}+\Delta^* d_{-\mathbf{k}\dn} d_{\mathbf{k}\up}
-|\Delta_\mathbf{k}|^2]
\eea
where $U>0$ is the on-site attractive interaction and $\Delta=\sum_\vec{k}\braket{d_{-\vec{k}\dn}d_{\vec{k}\up}}$ is the superconducting order parameter. As our major interest is in FF phase, we further introduce FF pairing in the Hamiltonian whose mean-field form is
\begin{equation} \label{eq:Hff}
H_{FF}= -U\sum_\mathbf{k} [\Delta_{\vec{q}} d^\dagger_{\vec{k}+\vec{q}\up} d^\dagger_{-\vec{k}+\vec{q}\dn}+\Delta_{\vec{q}}^* d_{-\mathbf{k}+\vec{q}\dn} d_{\mathbf{k}+\vec{q}\up}],
\end{equation}
where the superconducting order parameter is now defined as $\Delta_{\vec{q}}=\sum_\vec{k}\braket{d_{-\vec{k}+\vec{q}\dn}d_{\vec{k}+\vec{q}\up}}$, and the Cooper pair carries a finite momentum of $2\vec{q}$.
Then we construct Bogoliubov de Gennes (BdG) Hamiltonian whose matrix form is
\begin{equation} \label{eq:HBdG}
H_{BdG}=\sum_\vec{k} \Phi_{\vec{k}+\vec{q}}^\dagger
\begin{pmatrix}
\hat{H}_{M-TI}(\vec{k}) & \hat{H}_{pair}(\vec{q}) \\
\hat{H}_{pair}^\dagger(\vec{q}) & -\hat{H}_{M-TI}^*(-\vec{k}) \\
\end{pmatrix}
 \Phi_{\vec{k}+\vec{q}},
\end{equation}
where we define the 8-component spinor $\Phi_{\vec{k}+\vec{q}}=[\Psi_{\vec{k}+\vec{q}},\Psi^*_{-\vec{k}+\vec{q}}]^T$, and the pairing Hamiltonian is defined as
\begin{equation}
\hat{H}_{pair}(\vec{q})=
\begin{pmatrix}
U\Delta_\vec{q} i\sigma_y & 0 \\
0 & 0 \\
\end{pmatrix}.
\end{equation}
Alternatively, we may define
\begin{equation} \label{eq:HBdG2}
\hat{H}_{BdG}(\vec{k},\vec{q})=
\begin{pmatrix}
\hat{H}_{M-TI}(\vec{k}-\vec{q}) & \hat{H}_{pair}(0) \\
\hat{H}_{pair}^\dagger(0) & -\hat{H}_{M-TI}^*(-\vec{k}-\vec{q}) \\
\end{pmatrix},
\end{equation}
which satisfies
$H_{BdG}=\sum_\vec{k} \Phi_{\vec{k}}^\dagger \hat{H}_{BdG}(\vec{k},\vec{q})\Phi_{\vec{k}}$.
By setting $\vec{q}=0$, Eq. (\ref{eq:HBdG2}) becomes BdG Hamiltonian for BCS pairing.

With the Hamiltonian defined in Eq. (\ref{eq:HBdG2}), the BdG Hamiltonian can be diagonalized through Bogoliubov transformation as\cite{1966Gennes}
\begin{equation} \label{eq:dtog}
\begin{pmatrix}
d_{\vec{k}s} \\ d^\dagger_{-\vec{k}\bar{s}}
\end{pmatrix}
=
\sum_n
\begin{pmatrix}
u^*_{n,s} & v_{n,s} \\
-v^*_{n,\bar{s}} & u_{n,\bar{s}} \\
\end{pmatrix}
\begin{pmatrix}
\gamma_n \\ \gamma_n^\dagger
\end{pmatrix},
\end{equation}
where $s=\uparrow\; (\downarrow)$ is an index for up (down) spin, $\bar{s}$ indicates opposite spin index from $s$, $\gamma_n^\dagger$ ($\gamma_n$) is the creation operators for a quasi-particle (quasi-hole) operator, and $n$ is the eigenstate index. In Eq. (\ref{eq:dtog}), $u$ and $v$ are the matrix elements of the eigenvector matrix, $V$, which satisfies $\hat{H}_{BdG}V=VD$ where $D$ is a diagonal matrix containing $2n$ eigenvalues. Then, the correlation function at zero temperature is obtained as
\begin{equation}
F_{s\bar{s}}(\vec{k},\vec{q})=\braket{d_{-\vec{k}\bar{s}}d_{\vec{k}s}}=\sum_n u^*_{n,\bar{s}}v_{n,s}
\end{equation}
where the summation over $n$ is performed up to the filled states. The correlation function contains the information of the singlet order parameter, and it may be obtained from the following relation\cite{1966Gennes},
\begin{equation} \label{eq:Del}
\Delta_\vec{q}=\sum_\vec{k} \frac{1}{2}\left[
F_{\updownarrows}(\vec{k},\vec{q})-F_{\downuparrows}(\vec{k},\vec{q})
\right].
\end{equation}
The order parameter from Eq. (\ref{eq:Del}) is fed back to Eq. (\ref{eq:HBdG2}) until the change of each components of the density matrix reaches the convergence of $10^{-4}$. Finally, at zero temperature limit, the ground state energy is computed by\cite{1966Gennes}
\begin{equation} \label{eq:endif}
E_\vec{q}=\sum_n E_n + U\Delta^2_\vec{q},
\end{equation}
where we sum over all negative energies. Given the magnetization of the TI, we sweep over all possible $\vec{q}$ to obtain the minimum energy for finite $\vec{q}$ to determine the ground state energy of the FF pairing states, denoted by $E_{FF}$. In addition, we obtain BCS ground state energy by setting $\vec{q}=0$, denoted by $E_{BCS}$.
Then the favored supercodnucting ground state is determined by comparing $E_{FF}$ and $E_{BCS}$. It is important to note that we only add superconducting pairing interaction in the metallic region to model the proximity effect. As a consequence, the corresponding energetics is not dependent on the superconducting state of the TI. More precisely, we examine the FF superconducting state of the parent superconductor, even if the parent superconductor originally favors the BCS ground state. We look for the FF phase induced in the parent superconductor due to the spin orbit coupling from the IPE.


\begin{figure} 
\centering
\includegraphics[width=.5\textwidth]{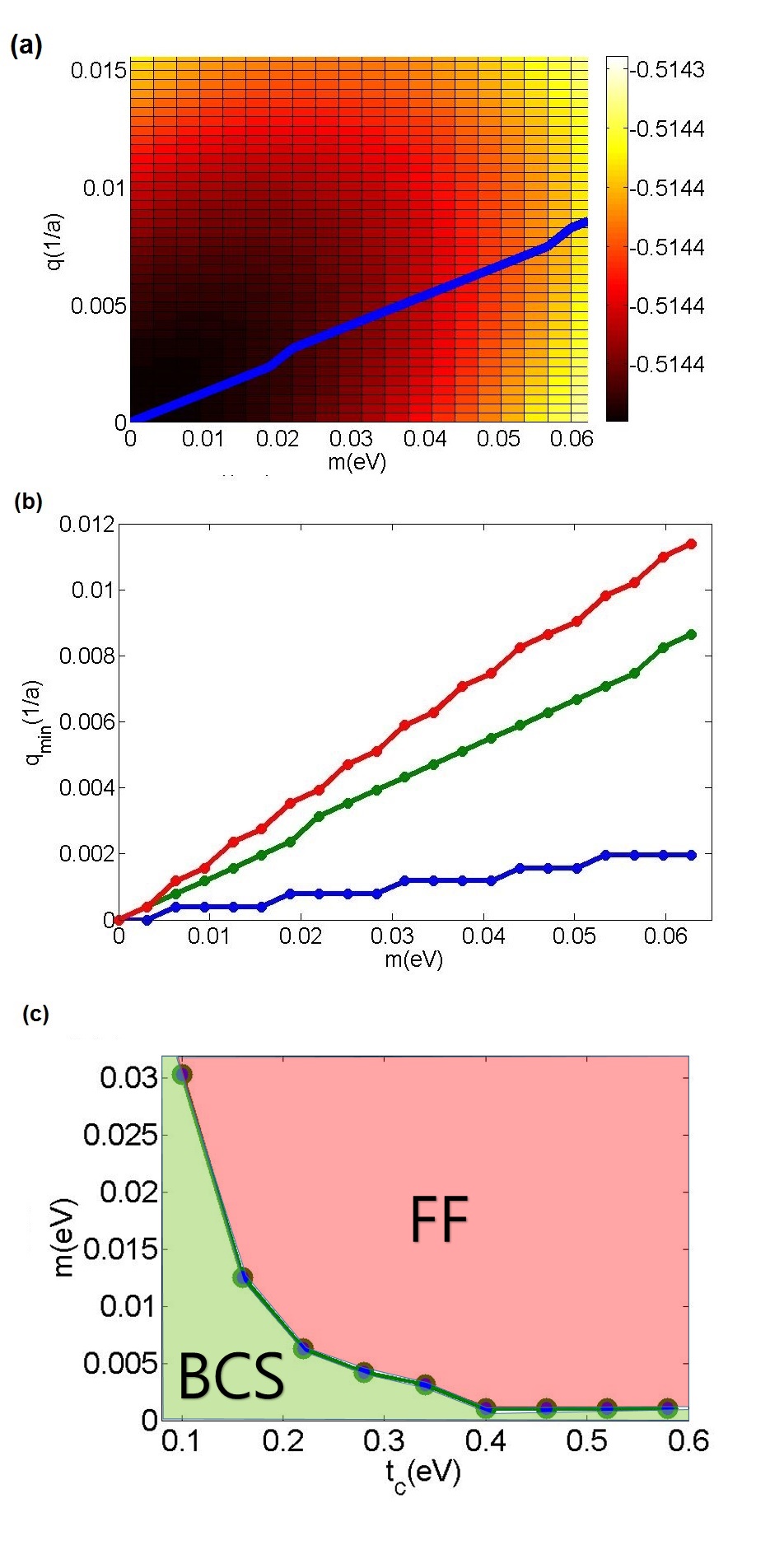}

\caption{\label{MQ} (a) The energy contour of the ground state energy as a function of $q$ and $m$. The region with $q=0$ corresponds to the BCS energy. As $m$ increases, we find the minimum of the energy occurs at non-zero $q$ which signals the FF ground state. The blue line shows the evolution of the location of the minimum as $m$ increases. We find that $q_{min}$ increases as $m$ increases. (b) The calculation of $q_{min}$ with various values of $t_c=0.3,0.6,1eV$. We find a clear linear dependence of $q_{min}$ respect to $m$. As $t_c$ increases the slope of the line increases due to the enhanced IPE. (c) By sweeping all possible value of $q_{min}$, we determine the pairing of the ground state with different values of $t_c$ and $m$. We find that the stronger $t_c$ and $m$ enhance the stability of the FF phase. }
\end{figure}

Using the self-consistent calculation of the superconductivity described in Eqs. (\ref{eq:HBdG}-\ref{eq:endif}), we now present the numerical calculations of the ground state energy with finite momentum pairing. Fig. \ref{MQ} (a) shows the energy contour plot as a function of the Zeeman field, $m$, and the momentum, $\vec{q}$, of the FF phase. As indicated by the blue line in Fig. \ref{MQ} (a), We find that the local minimum of the energy with the non-zero $\vec{q}$ exists at each Zeeman field, $m$. By tracking the location of the local minimum, $\vec{q}_{min}$, at each $m$, we find the linear relationship between $m$ and $\vec{q}_{min}$, which shows the clear signature that the finite Zeeman field in TI is inducing a FF ground states. This proximity induced local minima can be better understood by repeating the same calculations for various values of the coupling strength between the TI and the superconductor. Fig. \ref{MQ} (b) shows the dependence of $\vec{q}_{min}$ with the Zeeman field $m$ with various value of $t_c$. We find that the slope of the $\vec{q}_{min}$ increases as $t_c$ increases. The increase of the slope can be understood from the enhancement of the proximity effect due to the increase of $t_c$. From the calculation of $\vec{q}_{min}$, we conclude that the anisotropy of the Fermi surface due to the IPE favors the finite momentum pairing state. It is important to note that the clear linear dependence we find is limited in the weak Zeeman field limit. This linear dependence of the $\vec{q}_{min}$ has been similarly observed in other spin orbit coupled systems in the weak field limit\cite{Zheng2014}.

As we find that the energy of the FF state can be lower than the BCS energy ground state, we now draw the region within parameter space where the FF state is stabilized. This is calculated by comparing the ground state energy of the FF state with all possible momentum $\vec{q}$ and that of the BCS state. By computing the difference of the energies, Figs. \ref{MQ} (c) present the region of the parameter sets where the FF phase is favored. As we can see from the dependence of the Zeeman field, $m$, and the coupling strength, $t_c$, we find that the area of FF phase increases as $t_c$ increases and $m$ increases. This trend is a consequence of the stronger anisotropy of the metallic band resulting from the IPE with higher $t_c$ and $m$. Interestingly, in both weak and strong coupling regime of $t_c$, we still find the stable FF phase around 10meV strength of the Zeeman field, which is experimentally achievable value.

\subsection{Multiple-layers of the superconductor}


\begin{figure}
\centering
\includegraphics[width=.45\textwidth]{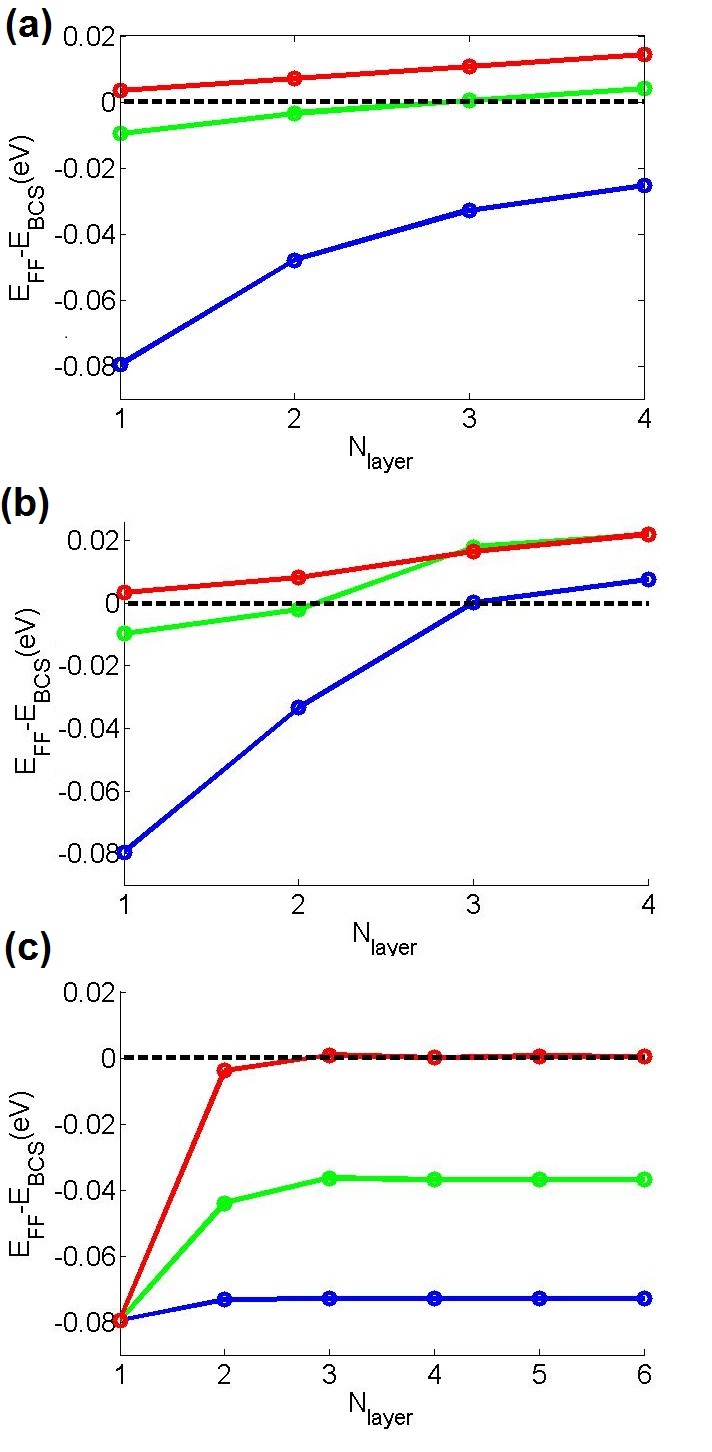}
\caption{\label{Qinall} The evolution of the ground state energy difference between the FF phase and the BCS phase. The negative value indicates the FF phase is more energetically favored. (a) The energy difference when $t_c>t_m=(0.1eV)$. We find that the energy gain of having the FF phase linearly decreases as $N_{Layer}$ increases. The different value of $t_c$ sets the initial energy gain when $N_{Layer}=1$. The red, the green and the blue indicates the value of $t_c=0.3,0.6,1eV$ respectively. (b) The same plot when $t_c<t_m=(1eV)$. The linear dependence disappears in the small $N_{Layer}$, however the same trend still holds in large $N_{Layer}$ (c) The calculation of the energy difference between the interface FF pairing and the BCS pairing. Unlike the homogeneous FF, we find that the energy difference saturates as $N_{Layer}$ increases, indicating that the interface FF might still survive in large $N_{Layer}$ limit. The red, the green and the blue indicates the value of $t_m=0.1,0.3,0.6eV$ respectively. } \label{fig:multi}
\end{figure}
In the previous section, we found the stable FF phase in the single layer limit of the superconductor. In this section, we now consider the case where the 2D superconductor is thick enough that the normal band has multiple Fermi surfaces. Unlike in the case of the single layer limit of 2D superconductor, the multi-layer superconductor may have a smaller region of the FF phase since the number of the bands, which originally favors the BCS superconductivity, is increased. Moreover, as shown in the section \ref{inverseproximity}, the IPE is short ranged effect in which the strength of the spin-orbit coupling exponentially decays from the interface between the TI and the superconductor. Hence, we expect the FF phase becomes more unstable as the thickness of the superconductor increases. In this section, we numerically calculate the thickness dependence of the stability of the FF phase. To do so, we consider multi-layer metallic Hamiltonian.
Using the metal-TI Hamiltonian we constructed in Eq. (\ref{eq:HMTI}), we introduce the Hamiltonian that connects two adjacent metallic systems, $\hat H_{m}=t_mI_2$, where $t_m$ is inter-layer hopping parameter. Then we construct the multi-layer metallic Hamiltonian from the following Hamiltonian construction:
\begin{equation}
\begin{split}
\hat{H}_{3D} =
\left(
\begin{matrix}
{\hat H_{M}} & {\hat H_m} & 0 & \cdots & \cdots & 0 \\
{\hat H^\dagger_m} & {\hat H_M} & {\hat H_m} & \cdots & \cdots & 0 \\
0 & {\hat H^\dagger_m} & {\hat H_M} & \ddots & & \vdots \\
\vdots & \vdots & \ddots & \ddots & {\hat H_m} & \vdots \\
\vdots & \vdots & & \hat H^\dagger_m & {\hat H_M} & {\hat H_{couple}} \\
0 & 0 & \cdots & \cdots & {\hat H^\dagger_{couple}} & {\hat H_{TI}}
\end{matrix}
\right).
\end{split}
\end{equation}
where $\hat H_{couple}=t_c I_2$ and $\hat H_M=t_mI_2$

By comparing the ground state energies of the Hamiltonians which have the FF and the BCS pairing over the entire region of the superconductor, Fig \ref{fig:multi} (a) shows the energy difference($\Delta E=E_{FF}-E_{BCS}$) between the FF and BCS as a function of the number of the superconductor layer, $N_{layer}$, where $t_c>t_m$. Although we expect $t_c<t_m$ regime is more experimentally relevant, this choice of parameters allows us to estimate the effect of the multiple superconducting layers using perturbation theory. Here we use the form of the FF order parameter that has the same momentum over the entire superconductor, which we refer it as the 'homogeneous FF'. We find that the energy gain of having the FF ground state quickly decays as $N_{layer}$ increases. Fig. \ref{fig:multi} (a) shows a steady decrease in $\Delta E$ and the full BCS pairing becomes more favored as $N_{layer}$ increases. The critical thickness where the FF and BCS ground state energy meet equal is also dependent on the coupling strength, $t_c$, between the topological insulator and the superconductor, and larger $t_c$ can sustain FF superconductivity in more metallic layers. This can be understood from the enhancement of the IPE when $t_c$ increases. Moreover, regardless of the value of $t_c$, we find that the same rate in the decrease of $\Delta E$ as $N_{layer}$ increases. This is due to the fact that the energy loss of having the FF pairing in the additional layers of the superconductor is simply proportional to the number of the layers. As a result, the energetic cost of having the FF pairing in the spin-orbit free superconducting layers increases with $N_{layer}$, where we expect a linear relationship between $\Delta E$ and $N_{layer}$. Furthermore, Fig. \ref{fig:multi} (b) shows $\Delta E$ when $t_c<t_m$. In this case, we cannot argue the multi-layer effect using the perturbation theory. Accordingly, we lose simple linear dependence of the energy as shown in Fig \ref{fig:multi} (a) when $N_{layer}$ is small. Nevertheless, the overall trend of decreasing $\Delta E$ as a function of $N_{layer}$ still holds.

In addition to the homogeneous FF order parameter over the entire region of the superconductor(homogeneous FF), we now postulate an additional form of the FF order parameter in which the finite momentum of the cooper pairs only survives near the interface region(interface FF). The inferface FF is defined as the order parameter profile with the momentum of the cooper pair $q$ that exists only within the first layer of the superconductor. The interface FF becomes more energetically favored than the homogeneous FF in the thick superconductor limit as the additional energy cost of having FF phase in the upper metallic region without spin-orbit coupling is no longer considered. This effect is numerically supported in Fig. \ref{fig:multi} (c). The Fig. \ref{fig:multi} (c) shows the energy difference as a function of the thickness and $t_m$, and we find that the energy difference saturates as $N_{layer}$ becomes larger than two. As the interface FF does not distinguish the upper layers of the superconductor from the BCS superconductivity, the resulting energy difference saturates with increasing superconducting layer. Interface becomes energetically advantageous as the homogeneous FF costs a constant amount of energy as the thickness of the superconductor increases. In addition to the thickness dependence, the interface FF has additional dependence on $t_m$. As we increase the values of $t_m$, the energetic difference proportionally increases in the interface FF. This is due to the increased Josephson energy between the interface and the upper layer of the superconductor. It is important to note that the ambiguity in the choice of the order parameter profile between the homogeneous FF and the interface FF is rather an artifact of the mean field approximation as one cannot determine the decaying length of the FF phase in a self-consistent manner.
%


\section{transport measurement} \label{transport}

\subsection{Four terminal Josephson junction}

\begin{figure}
\centering
\includegraphics[width=.45\textwidth]{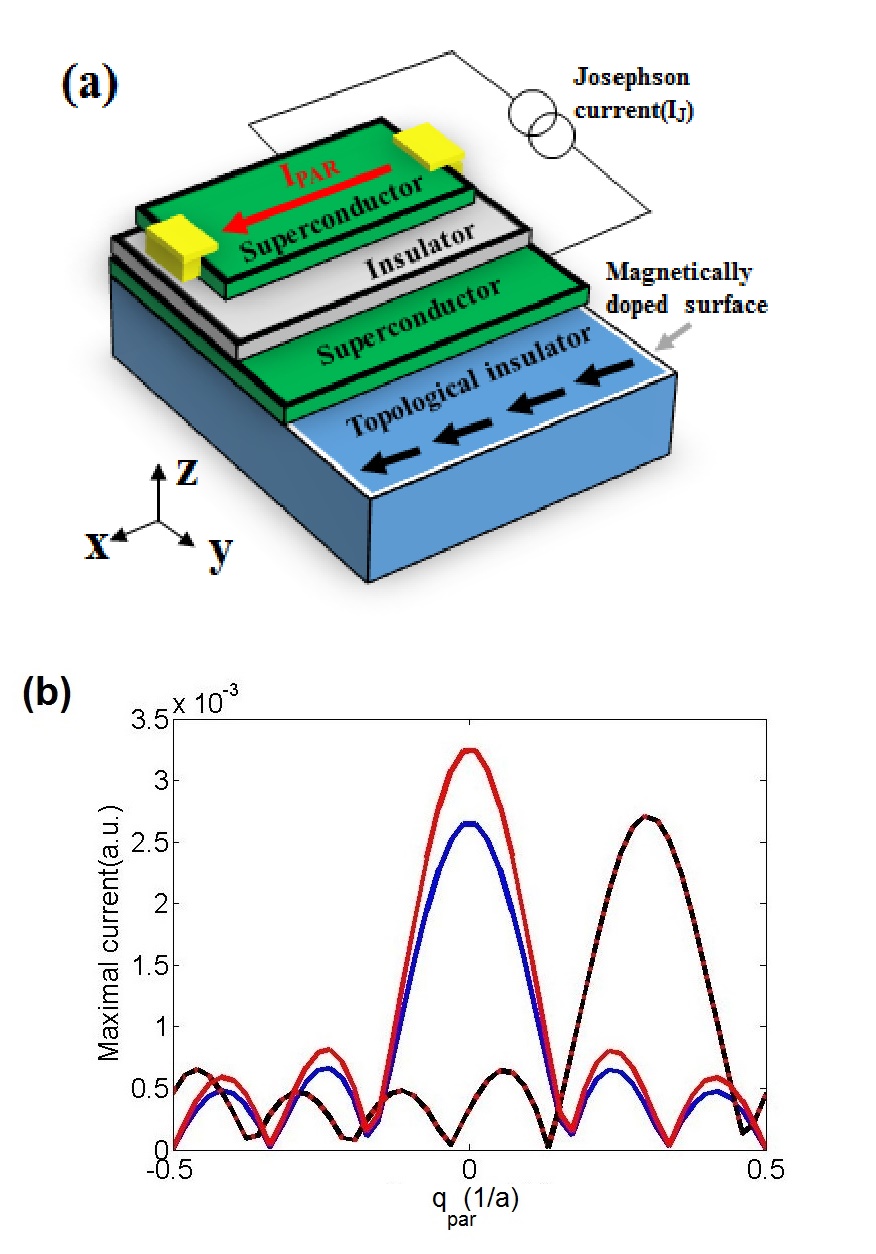}
\caption{\label{figtrans}(a) The schematic figure of the Josephson junction setup. On the top of the magnetically doped TI-superconductor junction, the normal insulator barrier is deposited, and the another superconductor is placed on the top of the normal insulator. The four terminal current is placed on the top of the superconductors. The two are attached on the different superconductors to drive the Josephson current. The other two are attached on the top BCS superconductor to drive the current in a direction parallel to the FF momentum. (b) Numerically calculated Josephson current as a function of the transverse momentum $q_{per}$. The blue, the black, and the red lines represent the Josephson current in the BCS pairing, the homogeneous FF pairing and the interface FF pairing respectively. We find that the blue(BCS) line has the maximum located at the $q_{per}=0$ and the black(homogenous FF) line has the maximum located at the $q_{per}=q=0.3$. The red(interface FF) line which has the peak at the $q_{per}=0$ are the interface FF phase with $N_{Layer}=2$.}
\end{figure}

In the previous sections, we analyzed the stability of the FF pairing. In this section, we now propose a Josephson junction transport and compare the transport signatures of the three different pairing scenarios: the conventional BCS phase, the homogeneous FF phase, and the interface FF phase. Fig \ref{figtrans} (a) shows the schematic figure of the transport configuration which consists of a Josephson junction between the TI-SC heterostructure and the conventional BCS superconductor separated by normal insulator. On the top of the superconductors we attach the four transport terminals. The two terminals are attached on the top of the two different superconductors so that the two junction can have a different phase of the superconducting order parameter by either applying the voltage bias or external current, $I_{J}$. The other two contacts are attached on the top of the BCS superconductor to drive the current in the perpendicular direction($I_{per}$) of the Josephson junction.

After establishing the setup of the Josephson junction, we now explain the manner in which current flows in this Josephson junction. Our setup utilizes the mismatch of the order parameter wave function on the interface between the BCS pairing and the FF pairing, this method has been similarly proposed to measure the LO state in the bulk doped inversion symmetric Weyl semi-metal\cite{PhysRevB.93.214511}. We first consider the weak coupling regime of the junction where the normal insulator is thick enough so that the Josephson current between the BCS and the FF superconductor can be approximated as,
\begin{gather}
\label{transporteq}
max(I_J)\approx t_{j} \int d^2 x \Delta_{top}(x)^* \Delta_{bottom}(x),
\\
\nonumber
I_J(\phi)\approx max(I_J)sin(\phi).
\end{gather}
where $t_j$ is the coupling strength between the junction. $\Delta_{top}$ and $\Delta_{bottom}$ is the order parameter wave function of the top and bottom superconductor respectively. The integration indicates the sum over the two dimensional junction region. As it can be seen from Eq. (\ref{transporteq}), the Josephson current is strongly suppressed when there exists a spatial interference pattern in the inner product of the order parameters of the two superconductor. As a result, the intrinsic spatial oscillations of the FF order parameter(i.e. $\Delta_{bottom}\approx |\Delta| e^{iqx}$) strongly suppress $I_{J}$ when it is coupled to BCS superconductor(i.e. $\Delta_{top}\approx |\Delta|$) in equilibrium. However, when $I_{per}$ is applied to the BCS superconductor, the BCS Cooper pairs possess the finite net momentum, resulting in the form of the order parameter, $\Delta_{top}=|\Delta| e^{iq_{per}x}$. The current induced spatial oscillations of the order parameter can cancel the oscillatory component of the FF order parameter in Eq. (\ref{transporteq}) when $q_{per}=q$, and recover $I_J$. Due to this momentum mismatch between the two superconductors, the Josephson junction between the FF state and the BCS state have a maximum of $max(I_J)$ under the non-zero parallel current, $I_{per}$, while the junction made with the two BCS superconductor always have a maximum in the absence of the parallel current.

We now illustrate this idea discussed above by numerically calculating the Josephson current. In this calculation, we model the normal insulator barrier using a small coupling strength $t_J$ between the superconductors. We also model $I_{per}$ by adding the finite momentum ,$q_{per}$, in the order parameter of the BCS superconductor. The Josephson current can be calculated from the full energy spectrum of the bound state in the junction by using the following formula\cite{1966Gennes}
\bea
J(\phi)=\frac{\partial E_{ground}(\phi)}{\partial \phi}
\eea
where $\phi$ is the phase difference between the two superconductors. $E_{ground}$ is the ground state energy. By explicitly sweeping $\phi$ from $0$ to $2\pi$, we derive the amplitude of the Josephson current as given in Eq. (\ref{transporteq}).
Fig. \ref{figtrans} (b) shows the amplitude of the numerically calculated Josephson current as a function of $I_{per}$ in the case of the three different scenarios of the superconducting order parameter. First of all, the blue curve shows the current in the case of the BCS pairing. As explained above, we find that the maximum of the current occurs in equilibrium when $q_{per}=0$ and the addition of the transverse current strongly suppresses the Josephson current as it introduces an additional spatial variation in the order parameter products. Unlike the case of the BCS superconductor, the black lines, which shows the Josephson current in the BCS-FF case, have maximum in the presence of non-zero parallel current which cancel the intrinsic spatial variation of the FF superconducting order parameter. As long as the FF state persists we find that this non-trivial Josephson current serves as an important signature which is distinguished from the conventional BCS pairing. Further, the red lines shows the transport of the interface FF pairing. Unlike the BCS and FF order parameter, we now find a crossover in the location of the maximum current layer increases. In the single layer limit, we find the maximum of the current occur in the same position as FF phase. However, as the $N_{Layer}$ increases more than two, we find that the current patter resembles the BCS phase, since the interface FF has identical order parameter to the BCS order parameter on the top. This shows that the Josephson current is only sensitive to the form of the order parameter near the junction region and the interface FF shows the distinct signature of the FF phase only in the thin superconductor limit.

\subsection{Y junction}

\begin{figure}[h!]
\centering
\includegraphics[scale=0.8]{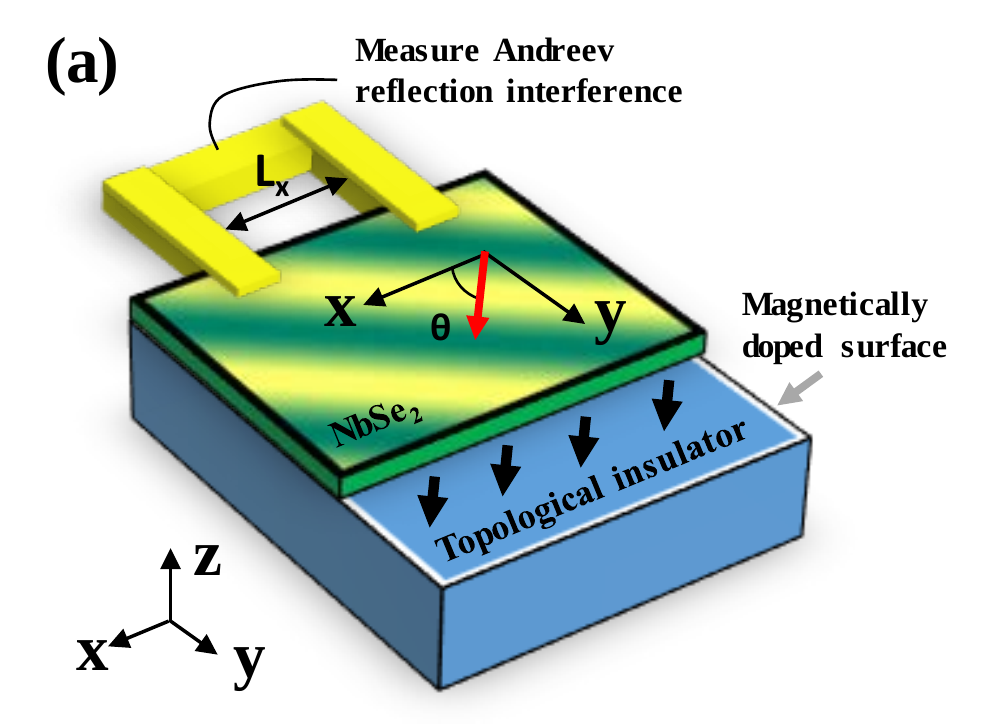}
\break
\includegraphics[scale=0.8]{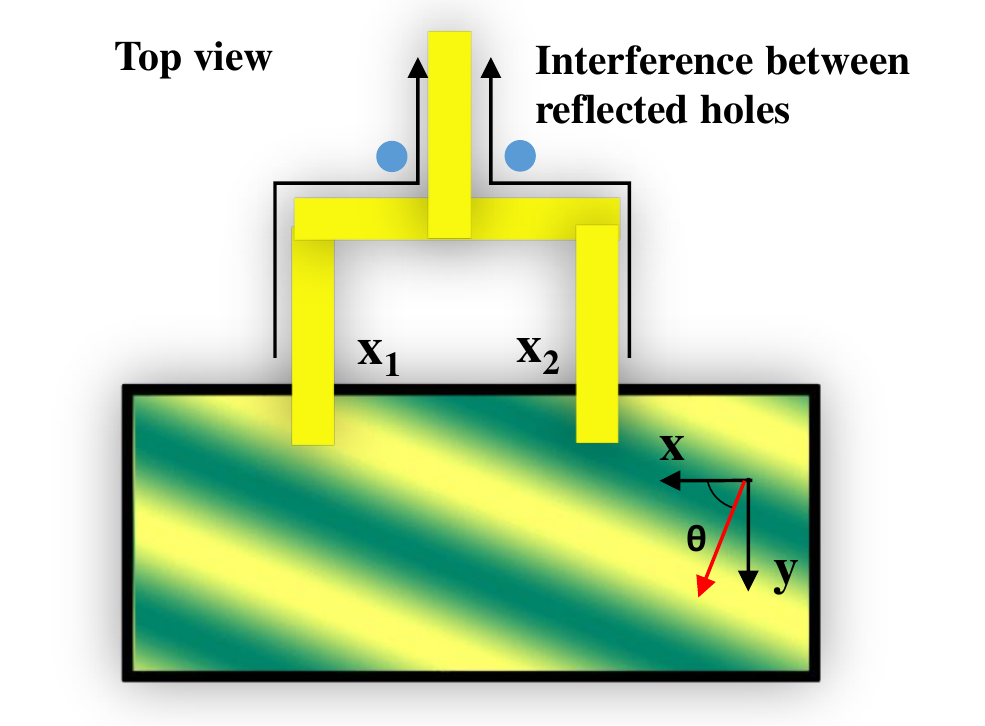}
\break
\includegraphics[scale=0.35]{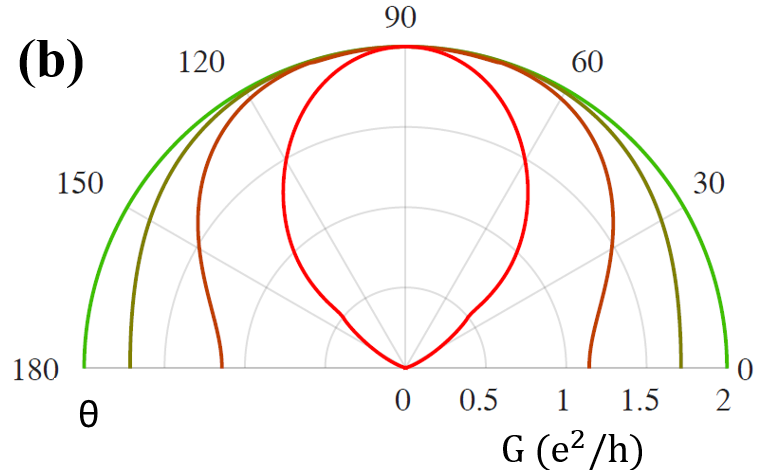}
\caption{(a) A schematic of the Andreev interferometer using Y-junction method to measure the unconventional superconductivity. (b) A plot of conductance as a function of finite momentum angle, $\theta$, of the FF phase. From the outermost to innermost, we plot the calculation results with different $\vec{q}$ where $|q|=0,\;\pi/8,\;\pi/6, and \;\pi/4$, respectively. We use a value of $v_F=6.61\times 10^{5}$ m/s for the Fermi velocity, that has been extracted from the metal Hamiltonian parameter, and plot the conductance at the incident electron the energy of $E=0.5\Delta$, where the $\Delta$ is superconducting gap. We set the barrier height at the interface of the metal arms and superconductor to be transparent.
} \label{fig:exp2}
\end{figure}

Another useful experimental method to detect FF phase is Andreev interferometer\cite{2014Xing}. Fig. \ref{fig:exp2} (a) shows the Andreev interferometer with the Y-junction with two arms separated by $L_x$ in $\hat{x}$ direction placed on the top of the superconductor. In the presence of the magnetization vector, $\vec{m}=|m|(sin\theta,cos\theta)$ on the TI surface, our analysis shows that the FF phase with the momentum vector, $\vec{q}=|q|(cos\theta,sin\theta)\perp \vec{m}$, is induced. In this case, the superconducting order parameter at each contact has different phases due to the phase modulation resulting from the finite longitudinal separation with respect to the momentum of the FF phase, $q$. We parameterize the different phases by assigning the order parameters $|\Delta| e^{-i\vec{q} x_1}$  and $|\Delta| e^{-i\vec{q} x_2}$ at the upper and the lower contacts respectively, where $x_1$ and $x_2$ are the coordinates of the upper and lower contacts. The phase difference between the two contacts is given as $\Delta \phi=q(x_1-x_2)cos\theta=|q|L_x cos\theta$. When the current flows through the Y-junction, the electrons injected from lower and the upper contacts undergo Andreev reflection process and reflected as holes. Due to the presence of the FF order parameter, the holes gain additional phases of either $\Delta e^{-iq_x x_1}$ or $\Delta e^{-iq_x x_2}$ depending on whether it is reflected from the upper or the lower contacts that comprise the Y-junction. The generation of this additional phase can be understood from an examination of the pairing Hamiltonian, $H_{pairing}(x)=\Delta e^{-iq_x x}c_x i\sigma_y c_x+h.c.$, at the interface between the contact, which annihilates a electron and create a hole with an additional phase of $\Delta e^{-iq_x x}$. Eventually, when the holes are collected to the central branch of the Y junction, the phase difference between different contacts generates an interference pattern as a function of $\Delta \phi\approx cos\theta$ and, most importantly, when $|q| L_xcos\theta=\pi$, destructive interference occurs and the conductance vanishes.

To illustrate the qualitative behavior of the Y-junction Andreev interferometer, we use the metallic Hamiltonian in Eq. (\ref{eq:Hm}) with assumed FF superconducting order. The conductance is obtained from Blonder-Tinkham-Klapwijk theory\cite{1982BTK} with an assumed interface barrier height that is transparent \cite{2014Xing}.The outermost line (Green solid line) in Fig. \ref{fig:exp2}b shows the conductance with no finite momentum in the superconducting system, or $\vec{q}=0$. The conductance shows a uniform distribution whereas we observe non-uniform conductance oscillation for $Q>0$. The innermost line (Red solid line) in Fig. \ref{fig:exp2} (b) shows $q_x=\pi/L_x$ where the phase difference between two arms is $q_x L_x=\pi\cos\theta$, and the conductance shows a destructive interference at $\theta=0$ and $ \pi$. Consequently, the signature of the conductance oscillation in Y-junction is a direct result of spatially varying nature of the order parameter. In addition, the Andreev interferometer is an optimal scheme for our proposal as one can adjust the angle of the finite momentum ($\theta$) before each transport measurements by applying in-plane magnetic field to adjust the orientation of the magnetic dopants rather than needing to fabricate different devices or multiple Y-junctions.
However, it is important to note that that the minimum momentum shift required to observe a clear destructive interference pattern is either $q=\pi/L_x$  or $L_x$ and that this quantity needs to be chosen within the scope of the maximum $Q$ that can be realized by the magnetic doping on the TI surface.

\section{conclusion} \label{conclusion}
In conclusion, we have studied the stability of the FF phase in magnetically doped TI-BCS superconductor heterostructures. We find that the FF state can be more energetically favorable than the traditional BCS pairing. This is due to the anisotropy of the Fermi surface in the superconductor that arises from the IPE where the normal band of the superconductor near the interface has the effective spin-orbit coupling and the Zeeman field. We find that the IPE quickly decays as the coupled state moves farther away from the interface into the bulk of the superconductor. As a consequence, the FF state gains more energy as the thickness of the superconductor increases and the stability of the FF state quickly decays. Nevertheless, in the thick superconductor limit, we find the FF phase can survive at the interface of the proximity structure.  We expect the FF pairing in our proposal can be experimentally measured through the four probe transport experiment utilizing a Josephson junction or through the Y-junction method.



\section{Acknowledgement}
M.J.P. appreciates David ChangMo Yang for helpful discussions. This work is supported by NSF CAREER ECCS-1351871.

\bibliography{reference}
\end{document}